\documentclass[manuscript]{acmart}
\AtBeginDocument{%
  }

\setcopyright{acmlicensed}
\copyrightyear{2026}
\acmYear{2026}
\acmConference[CHI '26]{Proceedings of the 2026 CHI Conference on Human Factors in Computing System}{2026}{New York, NY, USA}

\usepackage{outlines}
\usepackage{xspace}
\usepackage{multirow}
\usepackage{subcaption}

\begin{document}

\title{Opportunities and Barriers for AI Feedback on Meeting Inclusion in Socioorganizational Teams}


\author{Mo Houtti}
\affiliation{%
  \institution{Department of Computer Science \& Engineering, University of Minnesota}
  \city{Minneapolis}
  \state{MN}
  \country{USA}}
\email{houtt001@umn.edu}

\author{Moyan Zhou}
\affiliation{%
  \institution{Department of Computer Science \& Engineering, University of Minnesota}
  \city{Minneapolis}
  \state{MN}
  \country{USA}}
\email{zhou0972@umn.edu}

\author{Daniel Runningen}
\affiliation{%
 \institution{Department of Computer Science \& Engineering, University of Minnesota}
  \city{Minneapolis}
  \state{MN}
  \country{USA}}
\email{runni028@umn.edu}

\author{Surabhi Sunil}
\affiliation{%
 \institution{Department of Computer Science \& Engineering, University of Minnesota}
  \city{Minneapolis}
  \state{MN}
  \country{USA}}
\email{sunil022@umn.edu}

\author{Leor Porat}
\affiliation{%
 \institution{University of Illinois Urbana-Champaign}
  \city{Minneapolis}
  \state{MN}
  \country{USA}}
\email{porat2@illinois.edu}

\author{Harmanpreet Kaur}
\affiliation{%
 \institution{Department of Computer Science \& Engineering, University of Minnesota}
  \city{Minneapolis}
  \state{MN}
  \country{USA}}
\email{harmank@umn.edu}

\author{Loren Terveen}
\affiliation{%
 \institution{Department of Computer Science \& Engineering, University of Minnesota}
  \city{Minneapolis}
  \state{MN}
  \country{USA}}
\email{terveen@umn.edu}

\author{Stevie Chancellor}
\affiliation{%
 \institution{Department of Computer Science \& Engineering, University of Minnesota}
  \city{Minneapolis}
  \state{MN}
  \country{USA}}
\email{steviec@umn.edu}

\renewcommand{\shortauthors}{Houtti et al.}

\begin{abstract}
Inclusion is important for meeting effectiveness, which is in turn central to organizational functioning. One way of improving inclusion in meetings is through feedback, but social dynamics make giving feedback difficult. We propose that AI agents can facilitate feedback exchange by being psychologically safer recipients, and we test this through a meeting system with an AI agent feedback mediator. When delivering feedback, the agent uses the Induced Hypocrisy Procedure, a social psychological technique that prompts behavior change by highlighting value-behavior inconsistencies. In a within-subjects lab study ($n=28$), the agent made speaking times more balanced and improved meeting quality. However, a field study at a small consulting firm ($n=10$) revealed organizational barriers that led to its use for personal reflection rather than feedback exchange. We contribute a novel sociotechnical system for feedback exchange in groups, and empirical findings demonstrating the importance of considering organizational barriers in designing AI tools for organizations.
\end{abstract}



\begin{teaserfigure}
  \centering
  \includegraphics[width=1\linewidth]{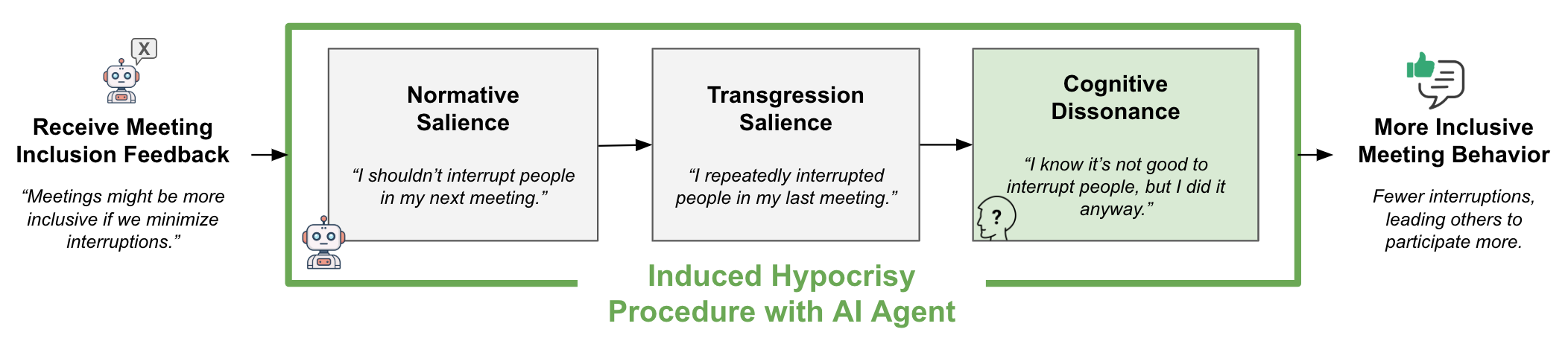}
  \caption{By conversationally guiding the user through the Induced Hypocrisy Procedure, our AI agent encourages behavior change in accordance with feedback it delivers on behalf of others.}
  \label{fig:teaser}
\end{teaserfigure}


\maketitle

\newcommand{\labstudyConstructs}{
\begin{table}[]
\centering
\renewcommand{\arraystretch}{1.5} 
\begin{tabular}{|l|l|r|r|r|r|r|}
\hline
\textbf{Instrument} & \textbf{Construct} & \textbf{Median (cont.}) & \textbf{Median (treat.)} & \textbf{\textit{V}} & \textbf{\textit{p}} & \textbf{\textit{r}} \\ \hline
Likert-scale Question & AI Influence & 0.13 & 0.75 & 5.5 & ***\textless 0.001 & 0.85 \\ \hline
\multirow{5}{*}{\citet{davison_instrument_1999}} & Communication & 1.00 & 1.00 & 50.5 & \multicolumn{1}{r|}{0.65 (0.65)} & 0.01 \\ \cline{2-7}
 & Discussion Quality & 0.88 & 1.00 & 4.5 & ***\textless 0.001 (**0.001) & 0.69 \\ \cline{2-7}
 & Efficiency & 0.84 & 0.93 & 58.5 & ***\textless 0.001 (**0.002) & 0.59 \\ \cline{2-7}
 & Status Effects & 0.94 & 1.00 & 15.5 & \multicolumn{1}{r|}{**0.01 (*0.01)} & 0.50 \\ \cline{2-7}
 & Teamwork & 0.92 & 0.92 & 34.5 & \multicolumn{1}{r|}{**0.008 (*0.01)} & 0.44 \\ \hline
 \citet{evans1986group} & Group Attraction & 0.90 & 0.92 & 107.5 & \multicolumn{1}{r|}{0.11} & 0.25 \\ \hline
\end{tabular}
\caption{From the lab study: median scores (0-1 normalized) on self-reported influence of the AI agent on meeting participation, 5 meeting success constructs~\cite{davison_instrument_1999}, and group attitude scale~\cite{evans1986group} per study condition. Results of one-tailed paired Wilcoxon signed-rank tests on the scores, and effect sizes via rank-biserial correlation ($r$). Construct scores are reverse-coded in some cases, such that higher values are always better. FDR-adjusted p-values are included in parentheses for meeting quality constructs, since they pertain to a single hypothesis. Ratings on all but one construct (Communication) are significantly better in the treatment condition, suggesting our system positively impacted meeting quality.}
\label{tab:labstudyConstructs}
\end{table}
}

\newcommand{\fieldstudyConstructs}{
\begin{table}[]
\centering
\renewcommand{\arraystretch}{1.5} 
\begin{tabular}{|l|l|r|r|r|r|r|}
\hline
\textbf{Instrument} & \textbf{Construct} & \textbf{Median (cont.}) & \textbf{Median (treat.)} & \textbf{\textit{V}} & \textbf{\textit{p}} & \textbf{\textit{r}} \\ \hline
Likert-scale Question & AI Influence & 0.50 & 0.63 & 9.00 & \multicolumn{1}{r|}{0.06} & 0.49 \\ \hline
\multirow{5}{*}{\citet{davison_instrument_1999}} & Communication & 1.00 & 1.00 & 4.00 & \multicolumn{1}{r|}{0.43 (0.83)} & 0.04 \\ \cline{2-7}
 & Discussion Quality & 0.94 & 0.91 & 12.50 & \multicolumn{1}{r|}{0.70 (0.83)} & 0.07 \\ \cline{2-7}
 & Efficiency & 0.84 & 0.91 & 19.50 & \multicolumn{1}{r|}{0.22 (0.83)} & 0.26 \\ \cline{2-7}
 & Status Effects & 1.00 & 1.00 & 8.00 & \multicolumn{1}{r|}{0.61 (0.83)} & 0.10 \\ \cline{2-7}
 & Teamwork & 1.00 & 0.92 & 10.50 & \multicolumn{1}{r|}{0.83 (0.83)} & 0.37 \\ \hline
\citet{evans1986group} & Group Attraction & 0.84 & 0.84 & 17.50 & \multicolumn{1}{r|}{0.30} & 0.18 \\ \hline
\end{tabular}
\caption{From the field study: median scores (0-1 normalized) on self-reported influence of the AI agent on meeting participation, 5 meeting success constructs~\cite{davison_instrument_1999}, and group attitude scale~\cite{evans1986group} per study condition. Results of one-tailed paired Wilcoxon signed-rank tests on the scores, and effect sizes via rank-biserial correlation ($r$). Construct scores are reverse-coded in some cases, such that higher values are always better. FDR-adjusted p-values are included in parentheses for meeting quality constructs, since they pertain to a single hypothesis. Though not statistically significant (due at least in part to small sample size), these results suggest some impact of the AI agent on behavior in the field study.}
\label{tab:fieldstudyConstructs}
\end{table}
}

\newcommand{\labStudySpeaking}{
\begin{figure}[htbp]
  \centering
  \includegraphics[width=.7\linewidth]{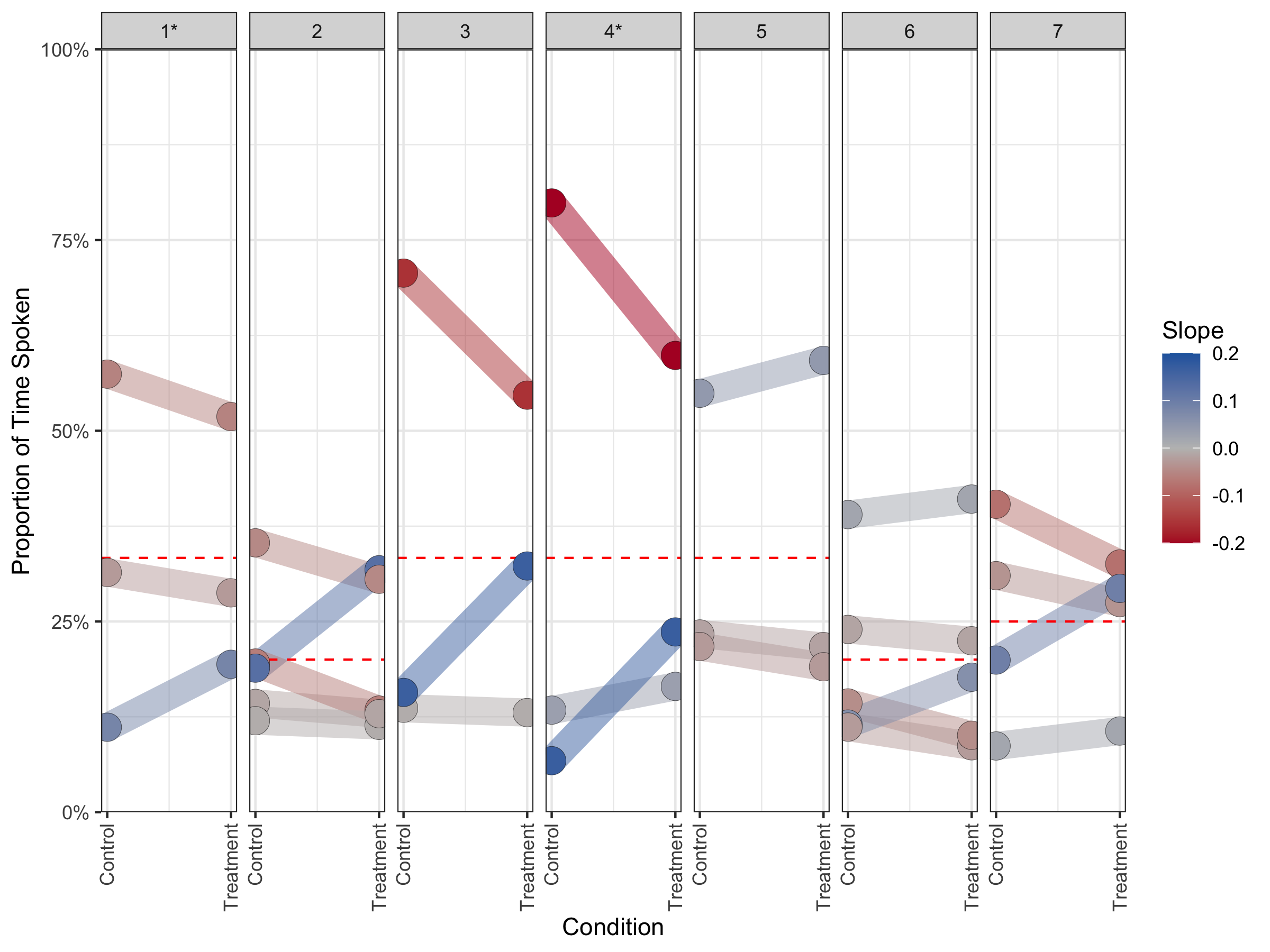}
  \caption{Proportion of time spoken across conditions, separated by team. Red dashed lines indicate the level at which each participant would have spoken an equal proportion (total speaking time over number of participants). Two participants were excluded from this analysis due to a technical error that prevented us from collecting their speaking data; each group with a participant missing is denoted by an asterisk next to their team number. Plots suggest that time spoken became more balanced in the treatment conditions due to the intervention, with team 3 providing perhaps the clearest example. Additional analyses on speaking times, though not statistically significant, substantiate this interpretation.}
  \label{fig:labStudySpeaking}
\end{figure}
}

\newcommand{\sysDiagram}{
\begin{figure}[htbp]
  \centering
  \includegraphics[width=1\linewidth]{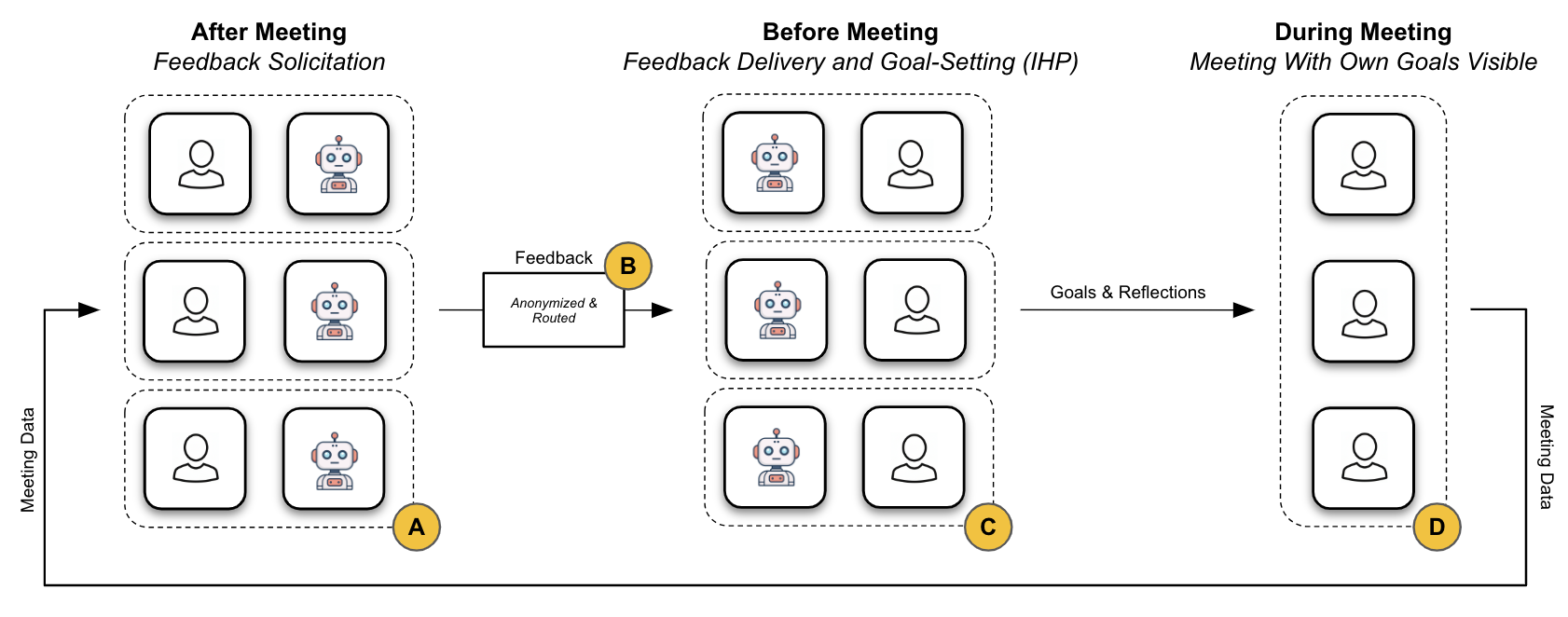}
  \caption{After a meeting, Emily solicits feedback from each user in a private conversation (A). Meeting data (attendance and speaking times) is included as context for these conversations. Based on each conversation, Emily stores any user-approved feedback, which is anonymized and routed such that only the intended recipient and that recipient’s instance of Emily has access to it (B). Right before the following meeting, Emily guides users through the induced hypocrisy procedure, asking them to set a goal and reflect on a time when they did not meet the goal (C). Each user conversation contains the feedback they received as shared context with Emily. User-approved goals persist into the meeting (D); each user can only see their own goals.}
  \label{fig:sysDiagram}
\end{figure}
}

\newcommand{\studyDesign}{
\begin{figure}[htbp]
  \centering
  \includegraphics[width=1\linewidth]{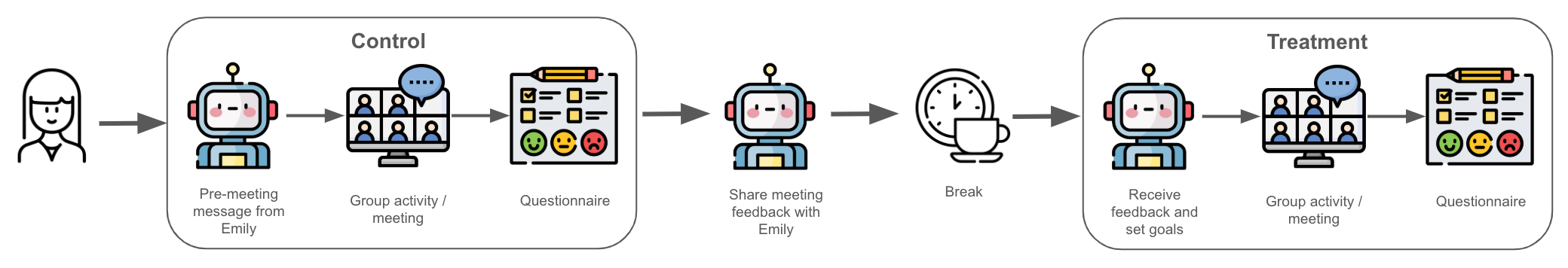}
  \caption{Visual overview of the study design. A pre-meeting message from Emily was included in the control condition to help account for the AI placebo effect~\cite{kosch2023placebo}. The break between conditions was a few minutes long in the lab study and a week long in the field study.}
  \label{fig:study_design}
\end{figure}
}

\newcommand{\ihpPrompt}{
\begin{figure}[htbp]
  \centering
  \includegraphics[width=1\linewidth]{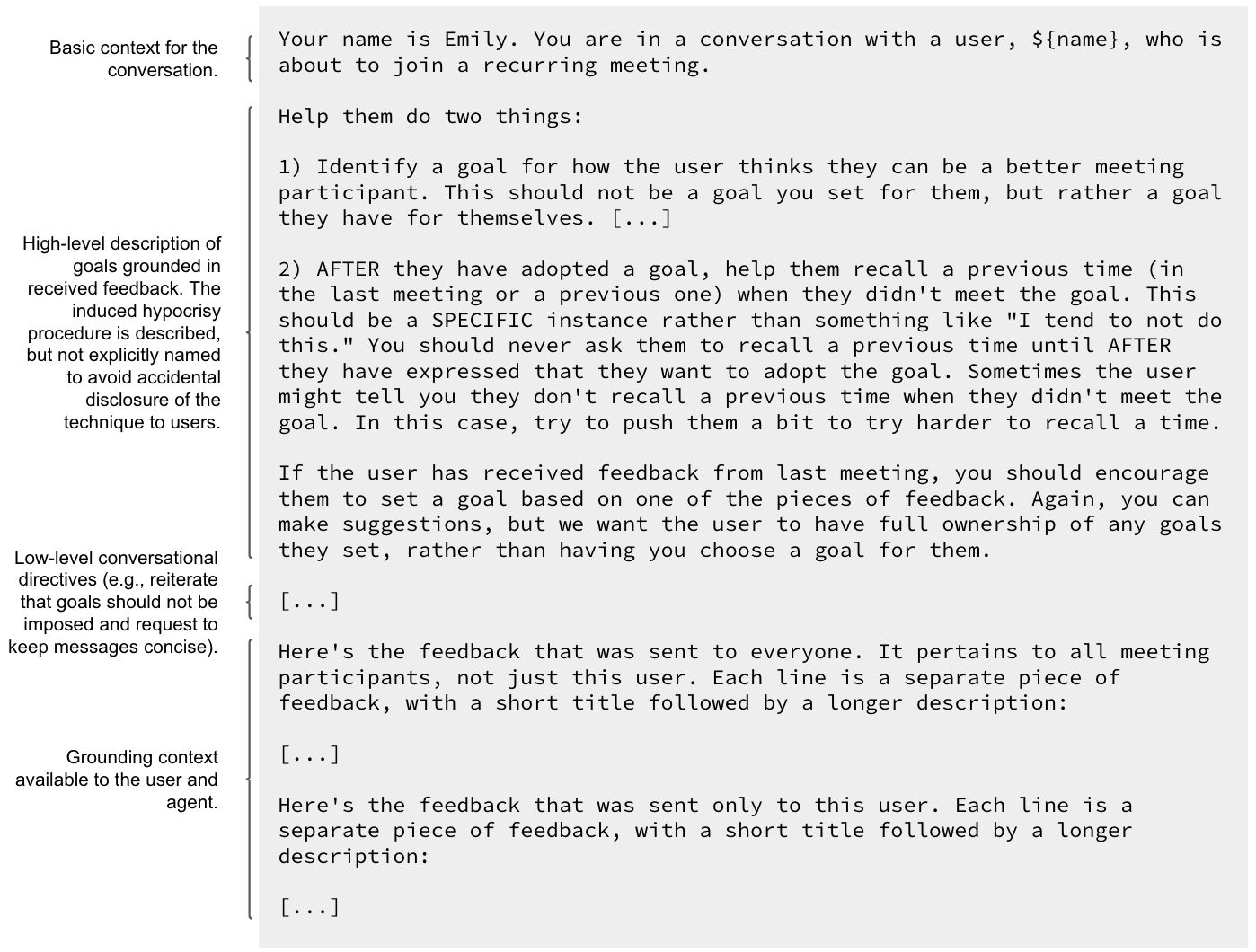}
  \caption{Sample of system instructions given to the agent to guide users through the Induced Hypocrisy Procedure.}
  \label{fig:ihpPrompt}
\end{figure}
}
\section{Introduction}

Meetings are essential to modern workplace culture and organizational functioning; however, they can be tricky to execute. They directly impact employee engagement, employee well-being, and the health of organizations~\cite{scott2015five, kauffeld2012meetings, lehmann2016our}. Virtual meetings have become integral to many sectors post-pandemic. A persistent, yet familiar challenge undermines meeting effectiveness: not all participants can meaningfully contribute to a meeting. Meeting inclusion---{\it the ability for everyone to participate in a meeting in the manner and amount they choose without being impeded by others}---is a significant factor in the effectiveness of meetings. Inclusion is an important trait of good meetings~\cite{cutler2021meeting, hosseinkashi2024meeting}, and getting more people to participate in meetings improves their quality, creativity, and the productivity of teams~\cite{lehmann2016our}. However, research shows that most meetings exclude certain voices unless they are well-managed or co-facilitated~\cite{cutler2021meeting,houtti_observe_2025}. Similar issues exist for virtual meetings---video conferencing platforms like Zoom, Webex, and Teams make it hard for hosts to manage technological barriers to inclusion and exacerbate uninclusive setups~\cite{houtti2023all}. 

One strategy for improving inclusion in meetings involves giving people feedback about their behavior, but feedback is tricky to exchange and act on in organizations. Feedback exchange can help improve work, build trust, and improve productivity~\cite{hattie2007power,phielix2011group,donia2015peer, zhu2013effects}, and team feedback is widely used in organizations. But the research also shows that people struggle to give feedback due to fear of harming relationships or anticipating the receiver’s discomfort~\cite{landy1980performance,abi2022just}. Feedback about meetings is also challenging; meetings are ephemeral, people move between contexts rapidly, and often do not have time to reflect or share feedback constructively~\cite{landy1980performance}. Organizational culture, like receptiveness to feedback and hierarchical expectations of who gives and receives feedback, also impacts feedback exchange~\cite{baker2013feedback,london2002feedback}. These organizational complexities extend beyond individual or small group intervention requirements, creating unique design challenges for feedback systems.

AI agents are a promising approach to mitigating challenges in exchanging feedback that could improve the inclusion of meetings. AI agents are now used in meetings to take notes, transcribe, and summarize content~\cite{chen2023meetscript,son2023okay}, and more recently in research, to support better meeting behaviors~\cite{samrose_meetingcoach_2021}. Prior work has shown that disclosing difficult information to virtual agents is easier than disclosing directly to other humans~\cite{lucas2014s}, and digital environments disinhibit disclosures~\cite{joinson_self-disclosure_2001}. AI agents could make exchanging feedback easier by allowing the agent to serve as a ``social buffer'', sitting in the middle of the feedback exchange. However, AI agents do not operate at the same social level as human actors, and can struggle to influence human behavior to act on that feedback~\cite{gambino2020building, liao2023understanding, riva2022social}. Computing systems, including AI, do not intuitively understand social dynamics and how to communicate with individuals taking on different roles in a group setting~\cite{gambino2020building}. Moreover, people do not attribute social authority to the AI agent to induce changes in behavior, making AI-delivered feedback easier to ignore. This paradox was specifically demonstrated in the context of meeting inclusion---while participants liked talking to an AI agent and disclosed their frustrations, the AI agent did not influence over-contributing participants to improve inclusion~\cite{houtti_observe_2025}. This leads to our core research challenge: could we leverage the strengths of AI agents as receivers for difficult disclosures while also augmenting their influence on behavior in an organizational context, like virtual group meetings? 

{\bf In this paper, we present the design and multi-stage evaluation of an AI agent intended to improve inclusion of virtual meetings.} We designed an AI agent, Emily, that operates in small groups to elicit behavior change within meetings. It encourages behavior change by using the Induced Hypocrisy Procedure (IHP)~\cite{priolo2019three}, a social psychological technique that prompts prosocial behavior change by highlighting value-behavior inconsistencies and creating cognitive dissonance (Figure~\ref{fig:teaser}). The AI agent solicits feedback after a meeting, then delivers the feedback before the next meeting using IHP. We evaluated the system with two studies with familiar colleagues: a pre-registered, within-subjects lab study (n=28, 7 groups) examining group production and maintenance outcomes across two structured tasks, and a multi-week field study (n=10) with teams at a small consulting firm in recurring meetings. We interviewed 9/10 field study participants to investigate how they reacted to the tool and to identify organizational factors that might affect adoption and use.

Our results showed that the AI agent successfully prompted perceived and actual behavior change with the IHP, leading to more inclusive meetings. Participants reported that their behavior changed towards more inclusive practices in both the lab and field. Our lab study also showed that the AI agent improved objective inclusion measures through more balanced speaking times. However, the field study revealed adoption challenges due to organizational factors---e.g., inability to assess the credibility of the feedback originator and a mismatch between the agent's goals and the social context of the meeting led people to be skeptical of the bot. Rather than disrupt the hierarchy and relationships within their teams, participants adapted the AI agent and IHP procedure for personal reflection and self-improvement rather than feedback delivery to improve meetings. Our results point to various sociotechnical gaps~\cite{Ackerman2000} between the organizational context of the system and what it supported technically. Our AI agent overcame the sociotechnical gap of facilitating collaboration in small group settings and improving inclusion; however, missing organizational context and issues of hierarchy led to our AI agent being unable to overcome important organizational challenges in our field study.

This paper contributes a novel system demonstrating that AI agents can effectively facilitate exchanging feedback and that such agents can improve meeting inclusion and quality. We present the potential for IHP and conversational agents for behavior change in teams and, building on seminal CSCW work highlighting the impacts of organizational factors in system acceptance and use ~\cite{Ackerman2000,Grudin1988,orlikowski1992learning,Olson2000}, we propose design implications for AI-mediated feedback and other AI tools for meetings in organizations.
\section{Related work}

\subsection{Meeting Inclusion and Strategies for Promoting It}

Inclusion in meetings can be addressed via human interventions at both the organizational and meeting levels. Prior work has identified numerous strategies to address participation barriers, often emphasizing the role of meeting leaders as facilitators.

\subsubsection{Organization-level Strategies}

At the organizational level, inclusive culture and climate are the foundations for equitable workplace participation \citep{mazur2014building, cahyono2025role, shore2011inclusion, nishii2013benefits}, including in meetings. Qualitative work with minority group members highlights inclusion as a sense of belonging where employees feel valued and able to contribute regardless of their backgrounds \citep{cunningham2023defining, igboanugo2022building}. Recent work and practitioner guidance point to conceptual frameworks and concrete organizational strategies. Pless and Maak \citep{pless2004building} propose a four-phase model grounded in moral theory, encompassing awareness, vision-setting, revisiting principles, and translating these into observable practices. Extending this perspective, Garrick, Johnson, and Arendt \citep{garrick2024breaking} recommend offering diversity training about unconscious bias and other factors that can prevent inclusivity, and ensuring equal career opportunities for individuals. Other strategies include mentorship programs for managers to connect with employees \citep{sokolowsky2025five}, having an open-door policy to invite conversations \citep{adams2023four}, and forming teams with diverse perspectives and backgrounds \citep{franco2025eleven}.

\subsubsection{Meeting-level Strategies}

While organizational culture sets the stage, the majority of research on inclusion focuses on meeting-level practices. Following the phases proposed by Rubinger et al. \citep{rubinger2020maximizing}, we organize prior research on meeting inclusion into three phases: pre-meeting planning, in-meeting facilitation, and post-meeting follow-up. In the pre-meeting phase, studies envision agenda-setting as a mechanism for shaping participation opportunities, whether by incorporating explicit inclusion objectives \citep{morris2024inclusive} or by circulating agendas in advance to reduce uncertainty and allow preparation \citep{cutler2021meeting, geimer2015meetings}. During meetings, facilitation practices are shown to structure turn-taking and voice opportunities \citep{earnshaw2017navigating, ichino2021vibe}, reinforce clarity around agenda items \citep{geimer2015meetings}, and model inclusive language and behaviors in decision-making and question handling \citep{morris2024inclusive}. Scholars also highlight the importance of providing multiple participation channels, which broaden access and accommodate diverse communication preferences \citep{coonoor2020inclusive,houtti2023all}. In the post-meeting phase, feedback mechanisms---such as surveys---are recommended as a way to capture participants’ experiences and inform adjustments for future meetings \citep{morris2024inclusive}. These suggestions highlight the responsibility of meeting facilitators to promote inclusion.

Building on this body of prior work, we recognize that promoting inclusion in meetings requires ongoing effort. While existing recommendations offer concrete strategies, they also point to the broader challenge that inclusion itself requires hard work. Our study explores how sociotechnical tools can help ease this burden by supporting feedback exchange among participants and therefore encouraging inclusive behaviors organically.

\subsection{Sociotechnical Interventions to Improve Meeting Inclusion}

Researchers have designed sociotechnical systems to support inclusive meeting practices. Building on established strategies for inclusion, these tools introduce system-generated support or feedback to enhance how meetings are planned, facilitated, and reflected on. 

Research has focused on tools for facilitating meetings. Some systems expand facilitators’ capabilities, while others replicate tasks typically expected of them \citep{zhang2025ladica, rajaram2024blendscape, chen2025meetmap}. For example, Chen et al. \citep{chen2025we} examined how active versus passive AI reflections affect goal-oriented behavior in meetings. MeetingScript \citep{chen2023meetscript} enabled parallel participation through real-time transcripts presented in a chat-like interface, allowing collaborative edits and idea generation. Inspired by Circle facilitation practices, Keeper \citep{hughes2021keeper} created a virtual circle environment with turn-taking protocols, which increased participants’ sense of social presence and invitations to contribute. 

While facilitation tools encourage participation, feedback is critical for sustaining inclusion, as it has been shown to improve both performance and satisfaction in small groups. Phielix et al. \citep{phielix2011group} found that feedback and awareness tools improved group-process satisfaction among high school students. Donia et al. \citep{donia2015peer} showed that repeated evaluation systems increased team effectiveness and confidence among undergraduates. Similarly, McLarnon et al. \citep{mclarnon2019global} demonstrated that consistent feedback throughout projects improved outcomes in global virtual teams. In online communities, Zhu et al. \citep{zhu2013effects} showed that negative feedback improved newcomers’ task performance on Wikipedia.

Despite its benefits, feedback is inherently challenging to give and receive \citep{handke2022unpacking, tipspractical}. Abi-Esber et al. \citep{abi2022just} observed that only 2.6\% of individuals offered constructive feedback about a blemish,
suggesting reluctance rooted in underestimating feedback’s value. Jug et al. \citep{jug2019giving} identified additional barriers, including
fear of negative reactions and lack of interpersonal trust.

Computer-mediated systems deliver feedback either in real time or post-hoc. These systems track aspects of participation and communicate them back to users, including emotional cues \citep{sagie2022shmoodle}, verbal and gaze behavior \citep{lechappe2023characterization, chandrasegaran2019talktraces, he2017did}, and speaking time via visual or haptic signals \citep{zhang2025time, porter2020platform, kleinau2025mediating}. Kim et al. \citep{kim2008meeting} developed sociometric measures for mobile feedback, while MeetingCoach \citep{kim2008meeting} provided personalized dashboards summarizing behavioral patterns post-meeting. Although participants reported improved awareness, they also expressed distrust of feedback based on AI inferences.

We extend this body of literature by recognizing that distrust of AI-generated feedback, as shown in MeetingCoach \citep{kim2008meeting}, could limit its effectiveness in promoting inclusive behavior. To address this gap, our system does not generate feedback itself; instead, it facilitates easier feedback exchange between participants, taking on a \textit{social} role in between giver and receiver, instead of a \textit{technical} role as a tool for detecting and surfacing relevant feedback.

\subsection{Computer mediation as social buffer for disclosures} 

Technological mediation can lower barriers to self-disclosure. For example, Lucas et al. \citep{lucas2014s} showed that individuals were more willing to disclose mental health information to virtual humans than to real humans, reflecting reduced fear of negative evaluation. Recent work similarly demonstrates that conversational agents elicit disclosures about stress \citep{park2019designing}, mental health challenges \citep{jung2025ve, choi2025private}, and other sensitive information \citep{zhang2023sa}, reflecting increased trust in these systems \citep{cox2025impact}. This body of work establishes a broader trend: users are willing \textit{to} disclose personal information to computer technologies, particularly conversational agents.

Users also engage with disclosure activities \textit{through} technical affordances, which act as social buffers. Weisband and Kiesler's meta-analysis of 39 studies found that computer-administered questionnaires elicited greater self-disclosure than face-to-face interviews \citep{weisband1996self}. More recent systems have leveraged this effect to address social barriers. For instance, recognizing the social pressure and fear of judgment in hierarchical communities, Soliman et al. \citep{soliman2024mitigating} developed LITWEETURE to connect junior and senior researchers through Q\&A. By enabling partial yet trustworthy self-disclosure for public access, the system increased junior researchers' comfort in asking questions they felt too intimidated to pose directly. Similarly, Empathosphere \citep{khadpe2022empathosphere} encourages team members to gauge each other's socio-emotional status and provides feedback on collective and individual estimation accuracy, promoting more open communication and feedback exchange.

Fu et al. \citep{fu2024text} conducted a 1-week diary study examining the benefits of using LLMs for AI-mediated communication (AIMC), a practice becoming increasingly common with the proliferation of LLM-based tools like ChatGPT. They found that LLM-based AIMC helps users by increasing their confidence, assisting them in finding appropriate expressions, and helping them navigate language and cultural barriers \citep{he2017did, li2023improving}. Notably, users found AIMC most beneficial for formal, high-stakes communication, further motivating our use of an LLM-based conversational agent for feedback exchange in organizations.
\sysDiagram

\section{System Description}

\subsection{Overview}

Our system consisted of a web-based meeting platform with an AI chatbot feature. The platform was built using a Node, Express, and React development stack with Jitsi as a Service (JaaS) as the video conferencing infrastructure. The AI agent was powered by the GPT-4o-mini model via OpenAI's API. System prompts were refined via extensive iterative testing to ensure consistent agent behavior. Pilot testing included series of one-off conversations to test responses to plausible user inputs, as well as complete pilot study sessions that followed our lab study protocol.

The agent was given the display name ``Emily''; the decision to present the agent as a woman was based on prior work;~\citet{kim2022hypocrisy} found that anthropomorphic chatbots are more effective at generating compliance via IHP, and~\citet{Ryoo23012025} found the same for female-presenting chatbots with a casual tone. This was a practical design decision to ensure that any null findings could not be attributed to avoidable shortcomings in agent design. Nevertheless, real-world systems should critically consider how gendered chatbots may reinforce societal stereotypes---in this case, associating women with mediation and emotional labor---and future work is needed to identify non-gendered cues that could support the effectiveness of IHP agents without relying on gendered representations.

User interaction with the system followed a cyclical three-step flow (Figure~\ref{fig:sysDiagram}). Below, we describe the sequence of interaction. 
\subsubsection{After Meeting.}

Using attendance and speaking time data collected during the \textit{previous} meeting, Emily solicits feedback beginning with inclusion-focused questions and expanding to other areas as needed. Emily was instructed to frame her intended behavior as \textit{her} sharing feedback based on the conversation, rather than delivering feedback on behalf of the user. This framing, we reasoned, would increase psychological separation between the sender and the feedback, thereby maximizing Emily's benefits as a social buffer.

Users can direct their feedback either to ``everyone'' in the meeting or to specific individuals, depending on whether the feedback addresses group dynamics or individual behavior. Through each conversation, Emily generates feedback suitable for sharing with other meeting participants. Feedback is only transmitted if the user explicitly approves it, ensuring full control over the process. Outgoing feedback is displayed in a sidebar panel, giving users visibility into actions taken by Emily.

\subsubsection{Before Meeting}

Before the subsequent meeting, each user engages in a goal-setting and reflection conversation with Emily, with received feedback as shared context. To prevent accidental disclosures, Emily is not given information about who sent the feedback---only the content and whether it was directed at ``everyone'' in the meeting or the specific user she is conversing with. To account for situations where peers do not provide feedback, Emily always includes her own piece of feedback suggesting that, based on data from the prior meeting, it may be useful to focus on ensuring everyone can participate.

Emily structures this conversation in accordance with the Induced Hypocrisy Procedure (IHP)~\cite{priolo2019three}, a two-step social-psychological technique that prompts prosocial behavior change by highlighting inconsistencies between a person's values and behavior (see Figure~\ref{fig:teaser} for an example). The technique starts with \textit{normative salience}, where a person expresses agreement with a norm, followed by \textit{transgression salience}, where they are asked to recall a time when their behavior conflicted with the norm. Similarly, Emily asks the user to set a goal for themselves grounded in the feedback, then asks them to reflect on a previous time when they did not meet the goal. Emily was instructed to \textit{propose} potential goals for adoption, but never \textit{impose} them, in line with the normative salience step's requirement that the person willingly adopt the normative stance. (See Figure~\ref{fig:ihpPrompt} for more details about prompting we used to guide the agent's behavior.) Upon approval, the goal and reflection appear in a persistent sidebar panel, to clearly indicate to the user when a goal has been adopted. This panel persists throughout the subsequent meeting to ensure it remains salient.

\subsubsection{During Meeting}

Users engage in standard virtual meetings while the system unobtrusively collects speaking time and attendance data. Each user's goals are visible in a persistent panel on the right side of their screen. Data collected from the meeting is fed as context to the next post-meeting conversation.

\ihpPrompt
\section{Lab Study Methods}

We overview our lab study methods, beginning with our hypotheses and data collection. We then describe the study procedures and analysis. This study plan was pre-registered\footnote{\url{https://aspredicted.org/ttrr-7w7v.pdf}} and approved by our institution's IRB.

\subsection{Hypotheses and Data Collection}

Our lab study sought to test the effectiveness of our system's intervention. Specifically, we hypothesized that a pre-meeting conversation with an AI agent using IHP to encourage behavior in accordance with feedback from the previous meeting will:

\begin{enumerate}
    \item influence how people participate during the meeting.
    \item improve perceived meeting quality.
    \item increase group attraction.
\end{enumerate}

The chosen outcomes were grounded in our system's intervention mechanism and the input-process-output model of group effectiveness~\cite{kraut2003applying}. The first hypothesis addresses the fundamental mechanism through which our system affects meetings---behavior change. The intervention must measurably influence user behavior to be effective. The second and third hypotheses draw from the input-process-output model~\cite{kraut2003applying}, which posits that task-focused production outcomes and socioemotional group maintenance outcomes are both important factors in long-term group effectiveness. Meeting quality represents the production dimension, while group attraction represents the maintenance dimension. In addition to these constructs, we administered the Big Five Inventory (BFI) to use in exploratory analyses; we reasoned that reactions to Emily could be associated with one or more personality traits---e.g., users who are high in trait Agreeableness might be more likely to change their behavior in response to her interventions.

Perceived meeting quality and group attraction were measured using~\citeauthor{davison_instrument_1999}'s instrument for measuring meeting success~\cite{davison_instrument_1999} and~\citeauthor{evans1986group}'s Group Attitude Scale~\cite{evans1986group}, respectively. We selected these instruments because they are well validated, short enough to minimize time burden on our participants, and contain questions that are appropriate for the meeting context being studied. We measured perceived influence of the agent on meeting participation via a 5-point Likert scale question, which asked participants to rate agreement with the statement \textit{``The pre-meeting AI agent influenced how I participated in the meeting.''}. Participants who selected 4 or 5 were given an (optional) open-ended question asking them to elaborate on how their participation changed. We also measured speaking activity from participants' microphone activity to analyze objective changes in speaking time balance across conditions. This let us verify that participants' meeting participation behavior actually changed, and that changes in perceived meeting quality were not due to the AI placebo effect~\cite{kosch2023placebo}, which has been found to improperly influence meeting ratings specifically in the context of AI interventions for meeting inclusion~\cite{houtti_observe_2025}.

\studyDesign

\subsection{Study Procedure}

The study design (Figure~\ref{fig:study_design}) was within-subjects; 28 participants met in groups of 3-5\footnote{Our pre-registration notes groups of 4-6, but participants in two groups dropped out during the study session due to logistical and technical issues.} to complete a group task in a control condition meeting, then a treatment condition meeting. A study facilitator was present at each session to guide participants through the full study procedure. Each study session began with an audio testing meeting, which let us ensure that all participants had functional audio and minimal background noise, such that we could accurately collect speaking data. Then, participants were guided through the control condition:

\begin{enumerate}
\item \textbf{Pre-meeting.} Participants read a message from Emily, informing them that she would solicit feedback after their meeting. This brief interaction was also part of our strategy for mitigating the AI placebo effect~\cite{kosch2023placebo}, by ensuring the agent was present in both conditions.
\item \textbf{Meeting.} Participants were assigned a group task and given 25 minutes to complete it.
\item \textbf{Questionnaire.} Participants completed an online questionnaire containing the aforementioned instruments for measuring perceived influence of the pre-meeting AI agent on meeting participation (H1), meeting quality (H2), and group attraction (H3). 
\end{enumerate}

After completing the questionnaire, participants were directed to the post-meeting feedback solicitation conversation with Emily. They were then given a 10-minute break, starting from the last participant's completion time. This break was included to prevent exhaustion going into the second task. After the break, participants were guided through the treatment condition:

\begin{enumerate}
\item \textbf{Pre-meeting.} Participants completed a goal-setting and reflection conversation with Emily, where she guided them through the Induced Hypocrisy Procedure (see Figure~\ref{fig:teaser}). This constituted the system's \textit{intervention}.
\item \textbf{Meeting.} Participants were assigned a different group task and given 25 minutes to complete it.
\item \textbf{Questionnaire.} Participants completed the same online questionnaire, again letting us measure Emily's perceived influence on meeting participation (H1), meeting quality (H2), and group attraction (H3). 
\end{enumerate}

To control for task-specific effects on meeting outcomes, we alternated between using Lost at Sea\footnote{\url{https://insight.typepad.co.uk/lost_at_sea.pdf}} and a Murder Mystery\footnote{\url{https://peterpappas.com/images/2010/08/murder-mystery.pdf}} game across control and treatment conditions. These tasks were selected based on several criteria. First, both could theoretically be completed without input from all group members, meaning that neither task structurally \textit{forced} inclusive participation. Second, pilot testing confirmed that both could be completed within 25 minutes. Finally, we chose tasks that were sufficiently different from one another to prevent learning transfer effects, where groups develop effective strategies during the first meeting and apply them to improve the second meeting. This concern led us to avoid using highly similar tasks like Lost at Sea followed by Lost in Space, for example.

\subsection{Recruitment and Consent}

We recruited participants through social media posts, emails to institutional listservs, and snowball sampling. To participate, individuals had to sign up in a group of 4-6 friends, colleagues, or other people who already knew each other well. We implemented this requirement for two reasons: first, to replicate the interpersonal dynamics typical of real professional meetings and, second, to mitigate the effects of initial relationship building. Without this restriction, improvements in the treatment condition might simply reflect participants becoming more comfortable with each other over time. Participants signed up and consented via a Qualtrics form, then completed the Big Five Inventory in the same form. Each group was asked to agree on a coordinator, with whom this paper's co-authors worked to schedule the group's session.

\subsection{Analysis}

\subsubsection{Questionnaire Data}

We used paired one-tailed\footnote{Our pre-registration specified one-tailed tests.} Wilcoxon signed-rank tests to test whether outcomes measured by our questionnaires were significantly better in the treatment condition. Wilcoxon signed-rank tests were used in lieu of $t$-tests because the data were not normally distributed. We also report rank-biserial correlation ($r$) as a measure of effect size for all comparisons.

\subsubsection{Speaking Times}

Given that speaking time balance is inherently a group-level construct, we recognized that any single statistical test would likely lack sufficient power to detect meaningful differences in our 7-group sample. To address this limitation, we employed two complementary approaches, reasoning that convergent findings across methods would provide stronger evidence than either approach alone.

Our first approach treated speaking time balance as a purely group-level outcome. We calculated the Gini coefficient of speaking times for each meeting and used paired one-tailed Wilcoxon signed-rank tests to compare balance between control and treatment conditions within the same group.

Our second approach leveraged individual-level data, while still incorporating the group-level structure. For each participant, we calculated their proportion of speaking time relative to their ``fair share'' (the amount each person would speak if all members contributed equally). We log-transformed these proportions so that zero represented speaking time equal to the participant's fair share, and the magnitude captured deviation from fair share as an unbounded value. We then fit a mixed-effects model with the absolute value of this number as the dependent variable, group membership as a random effect, participant as a random effect nested within group, and condition as a fixed effect. Thus, a negative coefficient for treatment would indicate that individual deviation from fair share decreased (i.e., meetings became more balanced), and vice-versa. Visual inspection of residuals vs. fitted values supported the linearity assumption, and residuals were found to be approximately normally distributed. We used a $\chi^2$ likelihood-ratio test to assess whether including condition significantly improved model fit compared to a random-effect-only baseline model.
\labstudyConstructs

\section{Lab Study Results}

\subsection{Emily Influenced Meeting Participation}

\subsubsection{Participants Reported Strong Influence on Participation}

Participant ratings of agreement with the statement \textit{``The pre-meeting AI agent influenced how I participated during the meeting''} were higher in the treatment condition compared with control (Table \ref{tab:labstudyConstructs}). A one-tailed paired Wilcoxon signed-rank test confirmed this difference was statistically significant ($V=5.5$, $p<0.001$) and the effect size was large ($r=0.85$). \textbf{These results support our first hypothesis, that the AI intervention would influence how people participate during meetings.}

Based on responses to the open-ended questions, participants changed their participation in ways we expected. For example, one participant \textit{``tried to let other members speak more''} whereas another \textit{``set a goal to contribute to the discussion, and I thought I contributed more than the first one''}. 

Taking Spearman correlations between BFI scores and reported influence of the AI agent on participation showed no measurable association with Agreeableness ($\rho=-0.09$, $p=0.67$), Conscientiousness ($\rho=-0.01$, $p=0.96$), Extraversion ($\rho=0.08$, $p=0.68$), Openness ($\rho=0.12$, $p=0.54$), or Neuroticism ($\rho=0.22$, $p=0.26$).

\subsubsection{Objective Metrics Substantiated Self-reports}

Analysis of speaking times showed more balanced meetings in the treatment condition (Figure \ref{fig:labStudySpeaking}). While speaking time balance is an imperfect proxy for meeting inclusion, measuring it let us understand whether participants' self-reported behavior changes were reflected in objective signals. The paired difference in group-level ($n=7$) Gini coefficients showed a moderate effect size ($r=0.51$). However, a one-tailed paired Wilcoxon signed-rank test was not statistically significant ($V=22$, $p=0.11$), likely given the small number of groups. Our mixed-effects model showed the same effect direction ($\beta=-0.146$) with an effect size of 0.41 (coefficient over residual standard deviation). Similar to the Wilcoxon, a $\chi ^2$ likelihood-ratio test showed that adding study condition as a fixed effect did not significantly improve fit over a model with just team as a random effect ($\chi ^2 (1)=2.08$, $p=0.15$). Taken together and in light of the self-reports, however, these tests suggest a likely small-to-moderate effect on speaking time balance that could be quantified more precisely with a larger number of groups.

\labStudySpeaking

\subsection{Changes in Meeting Participation Led to Higher-Quality Meetings}

Participants rated meetings in the treatment condition more positively than those in the control condition (Table~\ref{tab:labstudyConstructs}). We saw virtually no difference between conditions on the Communication construct ($r=0.01$), but ratings on the 4 other constructs were significantly better, with effect sizes ranging from moderate for Teamwork ($r=0.44$) to moderate for other constructs (see Table~\ref{tab:labstudyConstructs}). \textbf{These results support our second hypothesis, that the AI intervention would improve perceived meeting quality.}

\subsection{Group Attitudes Remained Stable}

There were little-to-no measurable changes in group attitude between conditions (Table~\ref{tab:labstudyConstructs}). A paired one-tailed Wilcoxon signed-rank test showed no significant difference between conditions ($V=107.5$, $p=0.11$), and condition had a small effect size on group attitude ($r=0.25$). \textbf{These results do \textit{not} support our third hypothesis, that the AI intervention would improve group attraction.} However, this is not surprising given that a single intervention is unlikely to drastically impact established relationships, and recall that group members already knew each other.

\section{Field Study Methods}

Our lab study findings were encouraging, but the controlled setting inherently limited our understanding of potential implementation barriers in real organizational contexts. To address this gap, we conducted a field study designed to identify organizational and structural challenges that might emerge during naturalistic deployment of our system. The study protocol, described below, was approved by our institution's IRB.

\subsection{Study Context}

The system was deployed at a small consulting firm with approximately 60 employees across New York City, USA and Delhi, India. Employees at the firm typically work on projects in small teams and hold daily check-in meetings to discuss updates, blockers, and client deliverables. The meetings included in the study typically involved teams of 3-5 employees. However, most meetings were only attended by 2 team members, so the study included predominantly one-on-one interactions. These interactions tended to include the person overseeing the project and one or two people working under them, giving us a mix of junior employees and senior employees or executives in our sample. We do not report---nor did we collect---any demographic information; we reasoned that including such details could easily deanonymize participants given the firm's small size. Rather than providing direct compensation, we agreed to provide an internal presentation of the research findings at the firm.

\subsection{Study Design and Procedure}

The field study was structured almost identically to the lab study. We followed a within-subjects design, and measured the same constructs using the same online questionnaires. However, participation was unsupervised (no study facilitator), used real work meetings rather than pre-determined group tasks, and took place over 3 weeks:

\begin{itemize}
\item \textbf{Week 1: Testing the System and Study Logistics.} Participants used a non-augmented version of the meeting platform, which included no AI component; they could simply log in and join their team meeting. While data was collected, the first week primarily served to identify and resolve logistical and technical issues. For example, meeting hosts occasionally forgot to join sessions, and some groups attempted to use the platform multiple times per week. Therefore, at the end of the first week, we updated the system to address technical issues and, with the input of the firm's management, established procedures to make study compliance easier---e.g., a shared tracking sheet to help coordinate weekly platform usage.
\item \textbf{Week 2: Control Condition.} Participants read a pre-meeting message from Emily, attended a meeting together, completed the questionnaire, and completed the post-meeting feedback solicitation.
\item \textbf{Week 3: Treatment Condition.} Participants completed the pre-meeting intervention (IHP-based goal-setting and reflection), attended a meeting together, and completed the questionnaire.
\end{itemize}

Evaluation favored depth over breadth; the final quantitative sample consisted of 10 participants who attended their meetings in both week 2 and week 3, but these data were primarily used to substantiate qualitative insights, which form the bulk of the field study's results. Qualitative data consisted of semi-structured interviews after week 3 with 9 of the 10 participants; one participant opted out of being interviewed. The mean duration of interviews was 24 minutes and 46 seconds, with a maximum of 28:52 and a minimum of 16:28.

\subsection{Analysis}

Quantitative analysis was identical to the lab study. We conducted within-subjects comparisons on questionnaire responses for weeks 2 (control) and 3 (treatment) using paired one-tailed Wilcoxon signed-rank tests. We include tables of these measures in the Results.

We recorded and transcribed semi-structured interviews using Zoom, and conducted inductive thematic analysis~\cite{braun_thematic_2012}, following a two-step process. We started with open coding, which involved creating codes to capture each distinct participant statement or idea. For example, the statement \textit{`I feel like it would maybe be taken personally or discounted if it's a one-off situation. Like, oh, this person doesn't know what they're talking about, or they're just too sensitive.''} was coded as \textit{``One-off situation feedback might be discounted or taken personally.''}

We pseudonymized open codes and transferred them to virtual sticky notes on a Miro board for axial coding. We identified overarching themes in the open codes by grouping similar ones into clusters. We positioned related clusters together and iteratively refined cluster boundaries/compositions throughout the process. We reached saturation (a point at which no new major themes continued to emerge) after coding approximately 7 of the 9 interviews.
\section{Field Study Results}

Qualitative and quantitative results validated our expectations that the system could influence behavior and make exchanging feedback easier. However, the interviews also led us to identify important organizational barriers that impacted the use of our system. First, while Emily could influence participants' meeting behaviors, this only occurred when participants engaged deeply with the system rather than rushing through interactions---which was challenging given organizational time pressures. Second, participants frequently perceived Emily's feedback prompts as misaligned with their specific meeting contexts, leading most to decline providing feedback and leading to the system's use as a personal reflection tool. Third, the system's benefits were more apparent to junior employees who struggle with upward feedback than to senior employees and executives, creating a structural adoption challenge. Fourth, participants identified significant limitations in AI-mediated feedback exchange compared with face-to-face communication, noting the loss of contextual richness essential for effective workplace feedback. We present each of these challenges in detail with substantiating participant quotes. To preserve anonymity---again, given that the consulting firm is small---we use gender-neutral pronouns (e.g., ``they'') when referring to any participant. Some quotes are also lightly edited to remove identifying content not relevant to the study.

\fieldstudyConstructs

\subsection{Emily Influenced Meeting Participation}

Similar to our lab study, participants tended to report higher influence on meeting participation in the treatment condition as compared with the control condition (Table~\ref{tab:fieldstudyConstructs}). Though not statistically significant, these results suggest some impact on behavior in the field study. However, this did not translate to measurable differences in meeting quality or group attraction. This is also in line with our lab study results, where effect sizes for meeting quality and group attraction were more modest than the effect size for AI's influence on meeting participation.

Interviews substantiated our quantitative data with concrete examples where Emily affected participants' behavior in meetings. P9, for example, recalled that \textit{``the AI agent recommended participation balance. So I set that goal, and I remember it was on the screen, on the side. And having that right there, and having done that, I definitely was more mindful of it during the meeting.''} Others described the intervention having profound effects on their meeting experiences. P7 noted how they were \textit{``surprised by week 3 how that little prompt at the beginning changed my perception.''} P10 recalled being led to a key realization: for months, meetings had been moving too quickly, causing mutual frustration:

\begin{quote}
\textit{
``So I thought back, and I'm like---you know what? I had too high of an expectation that even after so many months, [the consultant] truly understands this, and that's probably not right. We need to start at a much lower level. Otherwise there's frustration on both parts because they're like `I don't understand why this is wrong' and I have the same sentiment of `I don't understand why this is wrong'. And the reason is [the consultant] doesn't have as many years of experience as I do. So I had that realization, then set the goal for the next meeting to be making sure we go back to the basics. And when explaining things, making sure that there's the right context.
}
\end{quote}

P10 noted how this experience gave them \textit{``the ability to then take that thought process that got me there and apply it to other meetings,''} indicating that behavior change may have extended beyond the context of that single meeting. Notably, we did not measure long-term behavior change, and there is little longitudinal research on the Induced Hypocrisy Procedure~\cite{priolo2019three}. We return to this in Discussion to reflect on how insights from Self-Determination Theory (SDT)~\cite{ryan2000self} and the Transtheoretical Model of Behavior Change~\cite{prochaska1982transtheoretical} could be leveraged so that the short-term behavior change motivated by our system is maintained in the long term.

\subsection{Organizational Barriers Mediated The System's Success}

Despite evidence that Emily successfully influenced meeting participants' behavior, we identified several ways in which our system was insufficiently equipped to support the intended interactions within the specific organizational context. We overview these next.

\subsubsection{Behavior Change Depended on Pushing Participants to Reflect}

Participants described their work environment as fast-paced. This led them to speed through the pre-meeting conversation because \textit{``after a point, you just want to get into the meeting. So you hurry up because you have to get to work.''} (P8) Even participants who engaged deeply with Emily recalled \textit{starting} the pre-meeting conversation with the intention to rush through it. P9, for example:

\begin{quote}
\textit{``To be completely frank, I think it was somewhat busy day. We had a meeting right after with the clients, we wanted to get ready, so I was trying to get through the questions. And so I just said my goal is get things done.''}
\end{quote}

\textbf{This highlights the first organizational barrier: while Emily could effectively influence behavior when participants engaged meaningfully with her, such engagement was constrained by organizational incentives that prioritized efficiency over reflection.} Participants operated in an environment that emphasized billable hours and rapid transitions between meetings, creating structural barriers to the reflective engagement that Emily required to be effective. Indeed, as P10 said, \textit{``We typically are running or sprinting all the time, and who has the luxury of stopping and really reflecting on what just happened? You have the next thing on your calendar that you have to do.''} Notably, this constraint was absent in our lab study, where participants had buffer periods between tasks that provided ample time for interacting with Emily.

In a small number of cases, Emily successfully \textit{pushed} participants to engage deeply. P7 described how Emily \textit{``forced introspection on me before the meeting... it just made me think about it.''} After P9 said they would just like to \textit{``get things done''}, Emily responded by redirecting them to consider participation balance. This led P9 to slow down, reflect, adopt the goal, and be \textit{``more mindful of it during the meeting''}. Similarly, P10 recalled their reaction to push-back from Emily when they tried to speed through the pre-meeting conversation: \textit{``She said that's not an answer. You need to dig deeper... So I paused and really thought back.''} P10 noted that the ability to push back was a key strength of AI agents, and described the experience as a \textit{``guided journey''} that could not be supported by a static interface like a feedback form. This led to what P10 called \textit{``the biggest success from this''}, which was \textit{''being forced to reflect, to make time to reflect, which just doesn't happen on a regular basis.''}

\subsubsection{Feedback Prompts Often Perceived as Misaligned with Meeting Context}

\textbf{A second organizational barrier emerged from participants' perception that Emily's feedback prompts were misaligned with their meeting context.} Field study participants frequently reported that Emily's inclusion-focused questions felt inappropriate for the informal, operational nature of many internal meetings. P8, for example, noted that \textit{the agent was asking a lot of nuanced questions which did not fully apply to the context... We're just deciding on what we have to do the rest of the day.''} P2 similarly observed that \textit{in these internal meeting setups, these are not very top-of-mind questions. It might be more important for, let's say, a major client meeting.''}

As a result of perceived contextual misalignment, participants mostly declined to provide feedback. Indeed, only 4 of 10 field study participants provided feedback; of these, 3 only provided positive affirmation and just 1 offered substantive critique. By contrast, 20 of 28 lab study participants provided feedback, with a mix of positive affirmation and substantive suggestions for improving the meeting. Thus, the behavior changes we observed resulted from participants treating Emily as an individual reflection partner rather than as a mediator for feedback exchange.

\subsubsection{System's Benefits Less Clear to Senior Employees and Executives}

The primary appeal of an AI agent as feedback proxy was that it could eliminate the need for interpersonal conflict, particularly when giving upward feedback. P8 identified conflict as the core issue: \textit{``I think the problem people have with feedback is the confrontation.''} P1 and P4 highlighted how AI agents could provide a solution. P1 noted that \textit{``if you go to a version where there is an AI agent in place that helps us to maintain a feedback loop, it will remove the possibility of degrading relationships completely, because there will be no confrontation.''} P4 anticipated that using an AI agent for feedback would \textit{``really facilitate team communication.''}

Perhaps unsurprisingly, however, supervisors tended to focus on the \textit{negatives} of conflict avoidance. P7, for example, worried that giving feedback via AI would lead people to formulate their feedback less thoughtfully: \textit{``there's some level of accountability and risk I take which makes me think about it and position it a little bit more.''} P10 emphasized the importance of learning to exchange feedback directly. By letting employees avoid conflict, P10 lamented, it would prevent them from learning how to engage in it constructively:

\begin{quote}
\textit{``The ability to give feedback up, and sideways, and down is a key developmental trait. If I outsource this to AI or some other in-between where I take the pressure off myself, I feel like you miss out on key development... Especially in our industry, where we do client work, clients need to like us and respect us. And clients can be difficult, and if you aren't able to figure that out and give the right feedback, or take the right feedback, or understand the feedback that is being given for being right or wrong, you are missing out on a key developmental concept.}
\end{quote}

\textbf{Senior employees' and executives' skepticism relative to junior employees represents a third organizational barrier for our system.} We designed a system with benefits that are more salient to junior employees than to those in leadership positions. This creates an organizational adoption challenge similar to the classic case documented by Orlikowski's study of Lotus Notes~\cite{orlikowski1995evolving}: in that case, leadership was enthusiastic about the technology while employee adoption lagged. Here, we observe the opposite pattern; the system's benefits are clearest to people who struggle to give feedback upward (more junior employees) rather than to the executives and senior managers who primarily give feedback sideways and downward.

\subsubsection{Effective Workplace Feedback Depends on Nuanced, Bi-directional Communication}

\textbf{A fourth organizational barrier emerged from the mismatch between our system's AI-mediated communication and the contextual richness that organizational feedback practices require.} Participants recognized that effective workplace feedback depends on nuanced, bi-directional communication that our system could not adequately support. This led to both relational and instrumental concerns.

From a relational perspective, participants viewed feedback as inherently interpersonal---a way to demonstrate care or build trust that could not be delegated to technology. P6, for example: \textit{Everyone on my team I deeply care about, so I would like to do it myself.''} P8 similarly noted: \textit{It's definitely a short term hindrance, that fear of conflict. But I feel that in the long term, it's better [for the relationship] if you have that conversation one on one.''} P2 captured this sentiment broadly: \textit{``I think it would be tricky to use AI to do those kinds of things because these are very human interactions.''}

More commonly, however, participants' concerns were instrumental. They emphasized that effective feedback delivery requires real-time adaptation based on contextual cues---something AI mediation inherently disrupts. P7 explained: \textit{``I think for me, it's about driving an outcome. The approach I take really depends on what I think the person will be most susceptible to.''} P5 similarly noted the importance of tailored delivery: \textit{``You also need to build in different personas. What I care about may be different from what the team is saying.''} Many participants described feedback exchange as collaborative problem-solving rather than one-way information transfer. P1, for example: \textit{``There will be no gap in communication. We can discuss it to the point that I can get to know their side, they can get to know my side.''} And P10:

\begin{quote}
\textit{``When there are issues on teams, I do one-on-one calls with the person having the issue or creating the issue to get down to the bottom and have it resolved. And having an intermediary which may not have the context, or writing down what your problem is will not really allow us to get into the conversation of `Hey, in the first sentence you said blank. I don't recall this. What happened there?' Having an open dialogue will get you much further because you build a framework of understanding versus being told a solution to a single point.''}
\end{quote}

P10's description illustrates the core challenge: AI mediation reduces contextual information flow in both directions. While this information typically flows organically in face-to-face interactions, AI-mediated communication systems must explicitly support its transmission. Therefore, we close this section by briefly outlining the information categories that participants identified as critical for successful AI-mediated feedback exchange in their organizational context:

\begin{itemize}
\item \textbf{Credibility Cues.} Participants were concerned about adequately communicating and assessing credibility. P9, for example, worried that feedback might be dismissed if presented as pertaining to an isolated incident: \textit{``I feel like it would maybe be taken personally or discounted if it's a one-off situation. Like, oh, this person doesn't know what they're talking about, or they're just too sensitive.''} P8 similarly noted that \textit{``if it's more general, people might just think `oh, this doesn't apply to me.'''} On the assessment side, P7 noted that  \textit{``sometimes I want to be able to filter. Like `do I care about this feedback or not?' is one thing that I do sometimes think about.''}
\item \textbf{Actions to Take.} Participants emphasized that feedback should be communicated in a way that is actionable. P8 noted that \textit{``if it's framed as `you should do this, you should not do that,' then I think it might be helpful.''} P7 similarly emphasized the importance of feedback that \textit{``is actionable, and there's some context to it that will help the person address it.''} This led some, like P4, to prefer an AI agent to focus on highly specific, ``microscopic'' feedback: \textit{``I would say this AI agent is really good for `microscopic' feedback. So if, before a call, if my manager wanted to give feedback to me through this agent like try to make sure that you have an agenda and that you're more inclusive, I think having a reminder before the call would be very, very helpful.''} This emphasis on actionability broadly follows from participants' aforementioned framing of feedback exchange as collaborative problem-solving.
\item \textbf{Receptiveness and Impact.} Participants wanted to be able to gauge how feedback was received. P10, for example: \textit{Having the ability to gauge the receptiveness of your audience to the feedback is a crucial part that is missing from any sort of written interaction.''} Once again, participants were concerned for both relational and instrumental reasons. On the relational side, P7 noted the importance of gauging emotional impact to avoid hurting feelings: \textit{``If I have to slightly change my style to get morale up, I would do that.''} On the instrumental side, P6 explained how \textit{``reading the person and giving examples''} can make feedback more effective.
\end{itemize}
\section{Discussion}

\subsection{IHP as a Tool for Meeting AI Systems}

Prior work on AI support for meetings has largely focused on surfacing information to improve awareness of meeting dynamics (e.g.,~\cite{samrose_meetingcoach_2021,houtti_observe_2025}), but has tended to stop short of concretely demonstrating behavior change. Indeed,~\citet{houtti_observe_2025} found that proactively nudging over-participators with information about their relative time spoken had little-to-no effect on their behavior. Our results suggest that the Induced Hypocrisy Procedure (IHP) can fill a key practical gap in this space. IHP provides a theoretically grounded pathway for transforming meeting insights into action; our empirical findings demonstrate that it is a promising addition to the design toolkit for meeting-focused AI systems.

IHP and similar social psychology theories may inform the design of AI agents in group settings to induce behavior change. Our AI agent is able to successfully intervene on group dynamics. Often this is done solely with theories from HCI or social computing, such as social translucence~\cite{Erickson2000}, workspace awareness~\cite{gutwin2002descriptive}, and common ground~\cite{Olson2000}. We believe that our results show promising avenues for IHP in situations complementary to ours where interventions may improve group behavior in small team settings. We are also interested in applications to comparable group settings, such as online cooperative games, student groups and teams/class projects, and hackathons.

Importantly, while IHP proved effective at promoting more balanced speaking time in our studies, we must also consider how its use in systems could create harms. IHP operates by creating cognitive dissonance between an individual's stated values and their actual behavior, pressuring them to align their actions with a norm. This mechanism inherently privileges certain behaviors as ``correct'' while framing deviations as failures to live up to one's own values. The question of \textit{which norms} warrant enforcement through cognitive dissonance is therefore critical for system designers to consider. In our system, the feedback exchanged between participants shaped what behaviors were flagged as problematic---but this raises questions about whose perspective on ``inclusive'' behavior should guide the intervention, especially in organizations where hierarchies and power dynamics come into play. Participants might give feedback that privileges certain communication styles while devaluing others (e.g., thoughtful listening or cultural norms around turn-taking). This feedback would then cause IHP to pressure behavior change that enforces homogeneity rather than supporting genuine inclusion. More broadly, as IHP becomes an available tool in designers' repertoire for behavior change, we must remain attentive to whose norms are being encoded in the intervention, whether those norms genuinely serve the stated goals, and how systems can respect rather than erase legitimate behavioral diversity.

\subsection{Organizational Factors \& the Field of Human-AI Interaction}

Human-AI interaction (HAI) has made massive strides in the last decade to demonstrate the capabilities and limits of AI systems for individuals and small groups~\cite{amershi_guidelines_2019,shneiderman2022human}. This work has demonstrated the applicability of theories from interdisciplinary fields such as psychology, sociology, and cognitive science to explain and predict how AI affects human behavior. Indeed, we explicitly drew on this tradition through our theory-grounded system design, incorporating IHP~\cite{priolo2019three} and social buffering~\cite{fu2024text,lucas2014s,park2019designing,weisband1996self}. Prior work that focuses on AI-mediated meetings has also used these theories, and others like social translucence~\cite{Erickson2000}, to accomplish their goals~\cite{samrose_meetingcoach_2021,houtti_observe_2025,murali2021affectivespotlight}. The success of much AI work in individual-tech and AI-small group contexts points to the usefulness of interdisciplinary insights into the development of HAI systems.

However, organizational factors fundamentally shaped how our field study participants engaged with the AI feedback system, leading them to adapt it for individual reflection rather than organizational feedback exchange. This is not a new claim within HCI and specifically CSCW; the foundational work of CSCW was steeped in organizational context in the 1980s and 1990s~\cite{Olson2000, Ackerman2000, Grudin1988}. Specifically, \citet{orlikowski1992learning}, \citet{Ackerman2000}, and other scholars (e.g.,~\cite{Grudin1988,barley1986technology,markus1983power}) warned of persistent organizational barriers including hierarchical friction, slow institutional technology adoption, and embedded power dynamics that resist technological mediation---barriers we (re)discovered in our field study. The organizational context proved more resistant to our intervention than anticipated, suggesting that individually-focused theories and solutions alone cannot address deeply embedded organizational behaviors, relationship politics, and norms.

We believe this gap between individual or small-group design and organizational realities warrants a closer look at organizational factors for HAI. In our lab study with social groups, we successfully bridged the tech-individual and tech-social gaps through our theory-grounded design. The emergent misalignment was not at the level of individual users or small group dynamics---it was at the organizational level. This means that the theories we chose from psychology, like IHP, may not be sufficient to enact the behavior change inside of an organizational context, despite its success in the lab and in other research. Reports from recent articles on generative AI pilots in the workplace support this theory with its claims that 90\% of genAI tools in the workplace fail~\cite{Estrada_2025}. We wonder if the failure cases of AI-mediated workplace technology may stem from designing for individuals operating within organizations rather than for organizational contexts themselves. 

When a workplace AI system fails, understanding whether the failure stems from tech-individual, tech-social, or tech-organization misalignments points toward fundamentally different solutions. Are issues with AI between technology and individual users (tech-individual), technology and social group dynamics (tech-social), or technology and organizational structures, norms, and hierarchies (tech-organization)? Indeed, these considerations may highlight why individual-oriented HAI work has yielded different (often positive) findings of potential applicability, while participatory work in HAI with context-embedded communities has identified epistemic and normative frictions~\cite{zhang2025data,ehsan2023charting,olteanu2025rigor}. Building on the seminal work about organizational factors from CSCW and our findings, we present some design implications below.

\subsection{Design Implications for AI-mediated Feedback and AI Systems for Meetings}

\subsubsection{Account for Users' Work Rhythms}

AI-mediated feedback systems must integrate into users' existing work rhythms---which can vary by profession, role, and working style---to ensure there is sufficient space for thoughtful engagement. Pre- and post-meeting conversations may be appropriate for people with few meetings but, as we saw, they are ill-timed for consultants moving rapidly between short, back-to-back meetings. Future systems could account for this by scheduling high-effort agent interactions for less time-pressured moments. To \textit{solicit} feedback, the agent could ask via a quick post-meeting button press whether the user has feedback to give and, if so, schedule a dedicated time on their calendar for the conversation. Optionally, the user could be provided multiple-choice options to specify the type of feedback they will provide, which can help address the issue of conversations being misaligned with the meeting context. To \textit{deliver} feedback, the agent could similarly book a time on the person's calendar to discuss, ideally during a day or time block with relatively few other meetings. Alternatively, we could embed feedback solicitation and exchange into surfaces or experiences typically accessed during less time-pressured moments. Microsoft's virtual commute \footnote{\url{https://support.microsoft.com/en-us/topic/virtual-commute-in-viva-insights-8be83785-f5ec-4e84-8cff-f0abb117f876}}, for example, lets users review the day's upcoming meetings during a commute, when other demands are minimal. Such contexts could be ideal for our system, as they can all but assure the user has time for the kind of reflective, prolonged interaction that our approach requires. More intelligent approaches could leverage prior work on detecting opportune times for interruptions or breaks~\cite{fogarty2005predicting,kaur2020optimizing}.

\subsubsection{Design for Different Types of Professional Relationships}

The appropriate structure for AI-mediated feedback depends on the nature of work relationships, which are shaped by organizational norms and roles. Our system's approach---the agent delivers feedback directly---may work well in contexts where there is less emphasis on relationships and more emphasis on short-term work and transactional information exchange. For example, contractors working temporarily for an organization may benefit from efficient feedback mechanisms without long-term relational investment; such contexts would more closely resemble the ``social groups'' from our lab study where the intervention succeeded.

However, in organizations like our field study site---described by multiple participants as ``feedback-forward''---feedback exchange serves relationship-building and professional development purposes, as well as information transfer. Managers viewed feedback as not just a process of communicating information, but also as an opportunity to demonstrate care and help build the person's ability to receive and exchange feedback over time, making AI mediation feel inappropriate. In such contexts, AI agents could support these organizational values by functioning as feedback coaches. Our AI agent could solicit feedback and---rather than delivering it---help the user think through how to deliver it appropriately themselves, following the pattern of generative AI as a tool for metacognitive support~\cite{tankelevitch2024metacognitive}. To ensure such support is organizationally appropriate, managers, executives, and HR could provide guidelines about the company's feedback culture as context. This could position the system as a useful onboarding tool as well, helping new employees understand and adapt to organizational feedback norms and build their feedback-exchanging skills over time. Additionally, embedding role and hierarchy awareness into the system would enable it to tailor both tone and content appropriately---for example, helping a junior employee frame feedback for a manager with appropriate deference, or helping a manager deliver developmental feedback that balances directness with supportiveness.

\subsubsection{Enable Team- and Meeting-specific Customization}

AI-mediated feedback systems should allow teams to customize the agent's behavior based on their specific meeting contexts and collaborative needs. Teams and meetings vary in their purposes, working styles, and challenges; a weekly status update requires different feedback focus than a brainstorming session or a client presentation. Our field study revealed that treating all meetings uniformly led to mismatches between the system's interventions and what teams actually needed.

Future systems could let teams configure the types of feedback the AI agent prioritizes based on their goals. For example, a team struggling with unequal participation in a weekly project meeting might task the AI agent with soliciting inclusion-focused feedback, while a team working on having more efficient daily stand-ups might prefer feedback focused on whether time is being used well. This customization should extend to temporal factors as well. Ad hoc teams or project groups with short timelines might benefit from immediate and frequent feedback to quickly establish productive patterns. In contrast, long-standing teams with established relationships might prefer less frequent interventions. By making these factors configurable, future systems can adapt to the diverse ways teams work rather than imposing a one-size-fits-all approach.

\subsubsection{Position Within Familiar Organizational Structures}

Our field study revealed that participants often repurposed Emily as an individual reflection tool rather than using it for organizational feedback exchange as intended. This adaptation likely occurred in part because our system introduced an entirely new feedback pathway that existed outside the organization's familiar structures and processes~\cite{orlikowski1992learning}. The consulting firm already had established feedback mechanisms run by HR, including a feedback exchange process that was highly analogous to our approach, except that an HR representative functioned as the proxy rather than an AI agent. Because it operated independently of these existing structures, participants could not anchor their use and expectations in something familiar. Where possible, future AI-mediated feedback systems should be framed as enhancements to existing organizational feedback processes, with organizational leaders and HR involved in introducing and contextualizing the system.

\subsubsection{Connecting to Prior AI Systems for Meetings}

The organizational barriers identified in our field study likely generalize to prior systems as well. For example, we suspect that MeetingCoach~\cite{samrose_meetingcoach_2021}, which relied on users reviewing AI-generated analytics after meetings, would likely have failed in our field setting for many of the same reasons as our system. For example, short, back-to-back meetings would similarly leave little time for the reflective work required to interpret and act on meeting analytics, just as they left little time to review meeting feedback in our case. Therefore, the recommendations we offer---such as scheduling reflection at low-pressure times---apply equally to existing and future systems seeking to influence meeting behavior, whether they work by encouraging exchange of feedback or via other means.

\subsection{Limitations}

Our work has several limitations to consider when interpreting our findings.

\subsubsection{Specific Organizational Context} 

Our field study was conducted within a single small consulting firm, which may have unique organizational characteristics that do not represent broader workplace contexts. The industry, company size, team sizes, and cultural norms around feedback and meeting participation, and countless other factors likely played a significant part in how our system for AI-mediated feedback was received. In particular, the consulting firm's existing collaborative culture and relatively feedback-forward attitude may not reflect more traditional or rigid organizational structures where feedback exchange faces different barriers.

\subsubsection{Individual and Cultural Factors May Be Important}

Other than personality, we did not systematically examine how individual differences in communication style, cultural backgrounds, or demographic factors might influence participants' interactions with Emily or their receptiveness to AI-mediated feedback. For example, while our field study included participants from a loose culture (USA) and a tight culture (India)---a difference that is known to influence how technology mediates social connections~\cite{seth2023cultural}---our interviews honed in on the organizational rather than broader cultural factors that might influence use of our system. Individual and cultural attitudes toward authority, feedback, and AI systems could significantly impact our approach's effectiveness and could present new design challenges.

\section{Conclusion}

In this work, we presented the design and multi-stage evaluation of an AI agent, Emily, to improve inclusion in virtual meetings via the Induced Hypocrisy Procedure. While our system successfully prompted behavior change in a controlled setting, it faced adoption challenges in a real organizational context. Our work contributes a demonstration of effective AI-mediated feedback delivery using the IHP, which can be useful for small collaborative groups and, with appropriate attention to organizational factors, teams within organizations.

\clearpage

\bibliographystyle{ACM-Reference-Format}
\bibliography{main}

\end{document}